\documentclass[%
reprint,
superscriptaddress,
showpacs,showkeys,
 amsmath,amssymb,
 aps,
floatfix,
]{revtex4-1}

\usepackage{float}
\usepackage{comment}
\usepackage{graphicx}% Include figure files
\graphicspath{{Figures/}}
\usepackage{dcolumn}% Align table columns on decimal point
\usepackage{bm}% bold math
\usepackage[table]{xcolor}
\usepackage[colorlinks=true,citecolor=blue,linkcolor=red,urlcolor=violet]{hyperref}% add hypertext capabilities

\newcounter{reaction} 
\begin{document}
% You should use BibTeX and revtex.bst for references - no BibTex for ArXiv!!!!
%\bibliographystyle{revtex4}
\bibliographystyle{plain}
% marks overfull lines with blackboxes
\draft

\title{Potential Perturbation of the Ionosphere by Megaconstellations and Corresponding Artificial Re-entry Plasma Dust}
%Alternate: "The potential impact of megaconstellations on the magnetosphere"
% Optional argument for running titles on pages
%\title[]{}


% Authors, for metadata in PDF
\linespread{1.6}
\author{S. Solter-Hunt}

\vspace{5mm} %5mm vertical space

%\affiliation{Science Institute, University of Iceland,
%Dunhaga 3, IS-107 Reykjavik, Iceland}

\vspace{5mm} %5mm vertical space
\vspace{\baselineskip}

\date{\today}

\linespread{1}
\begin{abstract}
500,000 to 1 million satellites are expected in the next decades, primarily to build internet constellations called megaconstellations. These megaconstellations are disposable and will constantly re-enter and be replaced, hence creating a layer of conductive particulate. Here it will be shown that the mass of the conductive particles left behind from worldwide distribution of re-entry satellites is already billions of times greater than the mass of the Van Allen Belts. From a preliminary analysis, the Debye length in spaceflight regions is significantly higher than non-spaceflight regions according to CCMC ionosphere data. As the megaconstellations grow, the Debye length of the satellite particulate may exceed that of the cislunar environment and create a conductive layer around the earth worldwide. Thus, satellite reentries may create a global band of plasma dust with a charge higher than the rest of the magnetosphere. Therefore, perturbation of the magnetosphere from conductive satellites and their plasma dust layer should be expected and should be a field of intensive research. Human activity is not only impacting the atmosphere, it is clearly impacting the ionosphere. 

\end{abstract}

\maketitle

\section{Introduction}

The Van Allen Belts are two torus-like regions of trapped solar energizing particles that protect the atmosphere. The mass of the Van Allen Belts is 0.00018 kilograms. The masses of other parts of the magnetosphere (ring current, plasmasphere, etc) are not widely estimated but are less dense than the Van Allen Belts. The mass of one second generation Starlink satellite is 1250 kilograms\cite{Starlinkmass}, in which all of the mass will become conductive particulate in 5 years during re-entry demolishment and will be added to the lower ionosphere for an indefinite time. Thus, the space industry is adding enormous amounts of material to the magnetosphere in comparison to the magnetosphere's natural level of particulate and this is forming a layer of artificial plasma dust in the meteor ablation zone. Due to the conductive nature of the satellite material, this may perturb or change the magnetosphere. A depletion or change of the magnetosphere could subsequently have an impact on the atmosphere.

The magnetosphere is known to be weakening by 10-15\% \cite{mag} and satellites are typically involved in aiding that analysis. But in the 2020s and 2030s, satellites will become so numerous that they will form their own dynamic shell of conductive material. This shell of conductive material is now much greater than the radius of the earth. As satellites fill LEO (300 km) to GEO (36,000 km), this layer of conductive material extracted from the Earth's crust is 6 times the radius of earth. A comparison of space industry altitudes and regions with the Van Allen Belts is shown in Figure 1. Human spaceflight activity is thus creating vast regions of charged particles that may impact the Van Allen Belts and other parts of the magnetosphere in unknown ways. 

After consulting multiple magnetosphere models, the consensus was that it would take decades to simulate 500,000 satellites within a magnetosphere model and evaluate the potential impact. If the megaconstellations and their debris are perturbing the magnetosphere, their rate of growth is too fast for decades of simulations considering the potential urgent ramifications for the atmosphere. This may be a case where neither simulation or experiment can assess the question prior to the full deployment of 500,000 satellites. Thus, this must be considered based on overall planetary-level calculations between the magnetosphere and the megaconstellations. 

The space industry must fund more accurate chemical modeling of the atmosphere, ionosphere, and magnetosphere and the chemicals, debris, and materials they are adding to it. It is already suspected that the resulting re-entry alumina may increase ozone depletion \cite{Bolarticle}. Additionally, since space is not considered an earth environment, there is no regard for the sensitivity of the ionosphere, Van Allen Belts, plasmasphere, or magnetosphere. These plasma systems may indeed be more sensitive than suspected due to their low density. 

In addition to general mass calculations, ionosphere data and specifically the charge effectiveness or Debye length, is examined. It should be noted that this data is only possible because of in-situ satellite and rocket measurements and thus there is no pre-space-industry data available. 

\section{Method}
A simulation of the magnetosphere and the megaconstellations is not currently feasible, and the planetary-scale experiment is underway without a direct ability to diagnose the satellite-magnetosphere relationship because the satellites themselves detect the changes in the magnetic field. Thus, calculations on the mass, Debye length, and a small scale model are compared to gauge the issue. Debye length is studied by extracting the electron temperature and density from the Community Coordinated Modeling Center of the NASA Goddard Space Flight Center. Masses and other reentry statistics are studied from Jonathan's Space Report\cite{http://planet4589.org/space/con/star/stats.html}. As space-based development and pollution grow, there is a strong need for more multidisciplinary investigations and synthesis across materials, aerospace science, chemistry, plasma physics, and climate.

\section{Discussion}
\subsection{Mass}

The Starlink V2 satellite constellation (just one of many planned megaconstellations) intends to have 42,000 satellites \cite{http://planet4589.org/space/con/star/stats.html}, each the mass of a SUV, truck, or large car (1250+ kg). Each satellite has a planned lifetime of only 5 years (if successful and many satellites are failing sooner), thus in order to maintain the megaconstellation, 23 satellites per day will complete a re-entry burn in the upper atmosphere. This is approximately 26,308 kilograms (29 tons) of satellite re-entry material every day, just for the Starlink megaconstellation. Thus, every day, just for this one constellation, the conductive mass of material added to the upper atmosphere will be 150 million times greater than the mass of the protective Van Allen Belts. Thus, every second, just for this one megaconstellation, the mass of approximately 2,000 Van Allen Belts will be deposited into the ionosphere. Thus, the megaconstellations are creating their own plasma dust region vastly greater in mass than that of the Van Allen Belts or any other region of the magnetosphere. The South Atlantic Anomaly (SSA) is a region of the Van Allen Belts that comes closest to the Earth at 200 km and is avoided by satellites due to an increase in ionizing radiation. But the satellites themselves may be creating a more dangerous ionizing radiation zone with their reentry plasma dust. 

\begin{figure}
    \centering
    \frame{\includegraphics[scale=0.16]{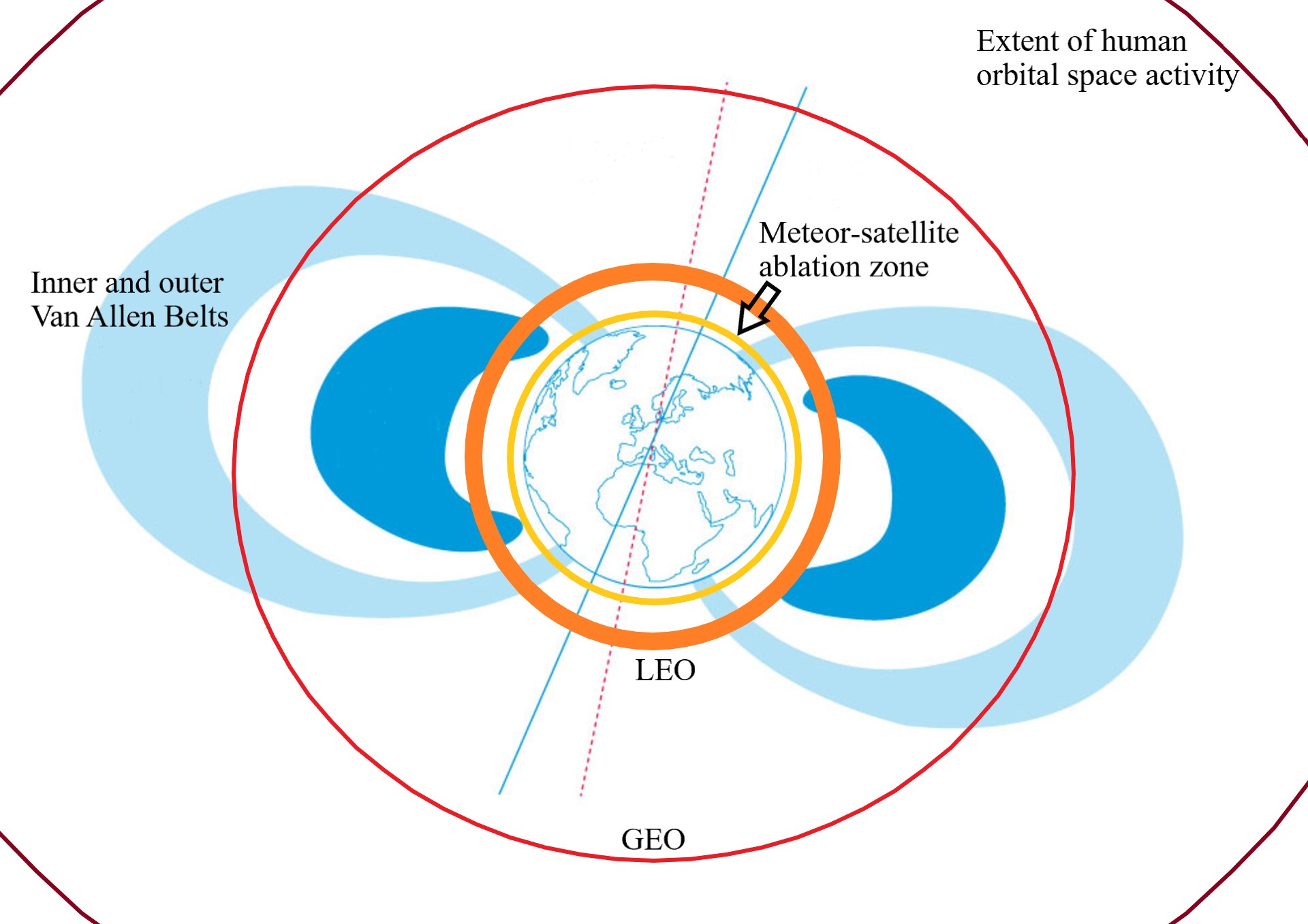}}
    \caption{Approximate schematic showing relation of human space activities to the Van Allen Belts. The relatively much-higher density debris and artificial plasma dust from human space activities are occurring under, through, and above the Van Allen Belts, suggesting that the much-higher density charged particles from these activities may disrupt the Van Allen Belts.}
    \label{fig:my_label}
\end{figure}

As of March 2023, 299 Starlink V1 (first generation) satellites have already re-entered \cite{http://planet4589.org/space/con/star/stats.html} at a mass of approximately 300 kg each. This amount of material is 500 million times the mass of the Van Allen Belts. In 2022 alone, the space industry polluted approximately 2 billion times the mass of Van Allen Belts (over 500 tons) in reentry particulate and material from all launches. Since the beginning of the space industry, approximately 20,000 tons of material \cite{http://planet4589.com/space/stats/reentry/data/remass.txt} have been demolished during reentry, meaning a similar amount may still remain as plasma dust. This amount is over 100 billion times greater than the Van Allen Belts. 

This re-entry material will be globally distributed since re-entries are globally distributed. The locations of re-entries are essentially randomly distributed. Space launch regions likely have greater densities of plasma dust development. This increase in ionization in these regions could have adverse impacts for launch success as ions can interfere with electronics.

The layer of growing re-entry particulate does not dissipate when the satellite demolishes. It stays stagnant in the upper atmosphere/lower ionosphere for several years before decay into the lower atmosphere in a best case scenario. In a worst case scenario, the particulate stays there indefinitely. Since this material is replenished every day, any potential natural decay to the ground may be negligible. It should be noted that rocket and satellite exhaust are also creating plasma dust \cite{Robarticle}. If the metal dust is settling into the atmosphere after several years, the repercussions of vast amounts of metal dust in the atmosphere includes dangers to the ozone \cite{Bolarticle} but any other potential impacts are unknown. 

%The satellites are primary composed of aluminium \cite{Bolarticle}. %Aluminium is known to block magnetic fields (REF). 

\subsection{Debye Length}
Even with 500,000 satellites, the average distance between those satellites will be in the hundreds of kilometers. However, the satellites will still have frequent near passes in the hundreds to thousands of feet, possibly overcoming the Debye length of the ionosphere and magnetosphere in some regions.  

More concerning is the density of aluminium particulate and other conductive metal particulate that is rapidly accumulating in the lower ionosphere after reentry demolishment of entire vehicles. These metals from satellites are already clearly measured in the stratosphere \cite{Metalarticle}. The satellites break up at 60 to 70 km, and here the ionosphere has a Debye length of approximately 1.5 meters on average between day and night conditions \cite{https://doi.org/10.1002/ctpp.201900005}. As shown in Figure 2, the Debye length goes up in the meteor ablation zone, however, this is likely from space debris and not meteors, as conductive material from satellites has far exceeded meteor material. If it is known that the Debye length has increased noticeably from meteor aluminium, then high concern should be placed upon the vast amounts of satellite aluminium. It is estimated that the meteor ablation zone gets approximately 50 tons of meteorite material per day - but the charged dust from that amount is approximately 1 percent (450 kg)\cite{Bolarticle}. Thus, one next generation Starlink (1250 kg) creates almost 3 times as much charged dust as all meteorite material in one day. If the amount of meteor aluminium and the corresponding charged dust has created an increase of Debye length by approximately 0.3 meters and satellite dust is approximately 60 times that (using the Starlink constellation alone), then assuming proportionality of this trend, the satellite aluminium from Starlink alone would potentially increase the Debye length by over 17 meters. This would be nearly double the Debye length of the Magnetosheath, which is 9.5 meters \cite{Debyeart}. The Debye length in the ionosphere and magnetosphere ranges from a few centimeters to a few tens of meters. Therefore, all satellite reentries are on a trajectory to create a band of charge higher than the rest of the magnetosphere. If this estimation is correct, Starlink reentries alone could be impacting the ionosphere.

\begin{figure}
    \centering
    \includegraphics[scale=0.47]{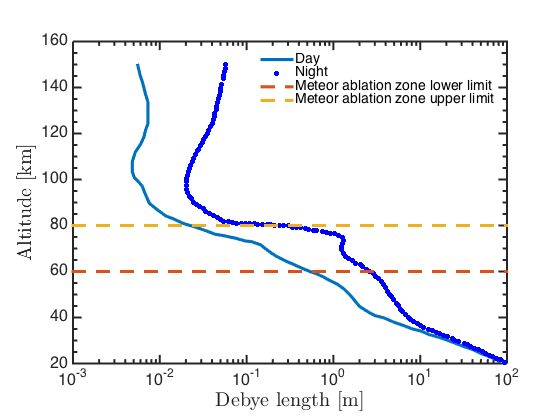}
    \caption{In this figure using CCMC data, it is shown that the electron Debye length increases at the meteor ablation zone in the lower ionosphere. This could potentially have been from space debris. This data is from 2010 and space debris has increased dramatically since. If it is believed that there is an increase in Debye length from meteorite aluminium, then the effect is stronger for reentry material, which has vastly exceeded the meteorite aluminium. As shown here, the increase in the Debye length is at the top of the ablation zone, further suggesting this is from space industry reentries.}
    \label{fig:my_label}
\end{figure}

The ion Debye length, which includes mobile ions, is given by 
\begin{align*}
  \lambda_D^2 & = \frac{\kappa k_B T}{4\pi q^2 c_0^2 k^2 }\\
\end{align*}
where $\kappa$ is the dielectric constant, q is the ion charge, $k_b T$ is the thermal energy, and $c_0$ is the ambient concentration of ions. 

The electron Debye length, which is extracted here from CCMC data, is given by 
\begin{align*}
  \lambda_D^2 & = \frac{\epsilon k_B T_e}{N_e e^2 }\\
\end{align*}
where $\epsilon$ is the permittivity of free space, $T_e$ is the electron temperature, and $N_e$ is the electron density. 

According ionospheric models, the lower regions of the ionosphere are particularly difficult for acquiring data and measurements. These regions are too low for satellite in situ measurements and difficult to sound from the ground typically because of the low ion/electron densities and high neutral densities. So data in this region on electron density relies on rocket measurements\cite{Ionoarticle}. However, data in a region slightly higher than the meteor ablation zone, already shows there may be concern because spaceflight regions are showing much higher Debye lengths than non-spaceflight regions, as shown in Figure 3. 

\begin{figure}
    \centering
    \includegraphics[scale=0.47]{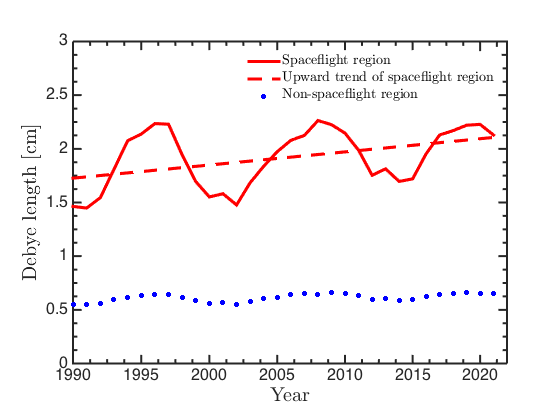}
    \caption{In this figure using CCMC data, it is shown that the electron Debye length at 100 km in spaceflight regions is much higher than non-spaceflight regions. The Debye length in spaceflight regions is trending up over time. This illustrates that human activity is having an impact on the ionosphere. Here Kennedy Space Flight Center is the spaceflight region and Reykjavik, Iceland is the non-spaceflight region. It is preferred to compare Debye length in the meteor ablation zone, but the exact values are more difficult to acquire because ionosondes and satellite measurements here are less reliable.}
    \label{fig:my_label}
\end{figure}

\subsection{Plasma drag}

Plasma drag occurs when charged particles are trapped in a planet's magnetic field. These charged particles can overtake the magnetosphere depending on their location and density. Drag is directly related to density and since it has been shown here that the masses of artificial plasma dust are going to create regions of charged particles with a much higher density than elsewhere in the magnetosphere, it should be expected that plasma drag within the meteor ablation zone is going to overtake the other regions of the magnetosphere. 

\subsection{Small-scale Scenario}

In a basic physics consideration, a spherical magnet (analog Earth) surrounded by a conductive spherical mesh (analog megaconstellations) would induce a drop in magnetic field outside of the conductive spherical mesh. This would allow solar energizing particles to more easily reach the satellite regions (and atmosphere) and cause satellite failures and thus further stratify the ionosphere with conductive particulate. The complication is that the spherical mesh in this scenario is very wide (6 times the radius of Earth) and whether this circumstance is more adverse is unknown.

However, the layer of the conductive particulate from satellite and spacecraft reentry demolishment may rapidly condense at current rates of activity. This layer of metal particles in the lower ionosphere, where the Debye length may become stronger than the natural ionosphere, will become a conductive shell of material. Inside the shell, the electric field tends toward 0. Thus outside the shell, the magnetic field tends toward 0. 

%Diagram
\subsection{Magnetosphere Loss on Mars and Dust on Earth}

It is thought that the loss of the magnetosphere on Mars may have involved separation of planetary materials in the core of Mars \cite{mars}. The space industry on Earth is taking vast amounts of conductive materials naturally found on the surface and in the crust and injecting them into the ionosphere and beyond, causing a new stratification of planetary material.

Mars, nearer to the asteroid belt, may have endured more meteor ablation involving aluminium and other metals. This may have played a part in the magnetosphere erosion, which the space industry is now accelerating with incessant satellite reentry demolishment. 

Additionally, further research regarding the Chicxulub impact \cite{DUSTarticle} indicates that the dust from the asteroid was a key mechanism in the extinction of dinosaurs and life in this period. The space industry is replicating this type of asteroid dust with more extreme chemicals, dynamics, mass, and combinations that have not had sufficient study at this time. At 1250 kg per satellite, only approximately 100,000 to 150,000 satellite re-entry demolishments are needed to meet the approximate mass of the Chicxulub impact. A new estimate \cite{Bolarticle} indicates that 1 million satellites are expected to be maintained and regularly demolished in total. 
\vspace{5mm} %5mm vertical space

\section{Conclusion}

With the estimated projected values for the Starlink megaconstellation alone (approximately one 1250 kg satellite per hour re-entering), every second the space industry is adding approximately 2,000 times more conductive material than mass of the Van Allen Belts into the ionosphere. It appears assumed by the space industry that the magnetosphere and Van Allen Belts are indestructible, when in fact the nature of their composition is delicate and billions of times less massive than the conductive material being added to the magnetosphere on a regular basis. Continued monitoring and analysis of this artificial charged dust in comparison to the natural charged dust is needed.  %Perhaps the Van Allen Belts should be tested for their strength and the potential density of aluminium satellite particulate to confirm their potential interference. 

It is known that man-made chemicals can endanger the atmosphere, and the megaconstellations are not just depositing dangerous chemicals, they are depositing huge masses of conductive material. It is known that injecting fossil fuels into the atmosphere, originally from below the Earth's crust, is risking the habitability of Earth. It appears likewise for injecting metals from the ground and crust into the ionosphere. Using just estimates from Starlink reentry masses, the Debye length of the ablation zone may increase to over 17 meters if the planned reentry demolishment of 23 satellites per day occurs. This amount, if globally spread and amplified by hundreds of other planned and developing megaconstellations, would cause perturbation to the magnetosphere because it would out-compete the Debye length elsewhere in the nearby magnetosphere and magnetosheath. It may be the case that satellites, reentry particulate, and conductive space debris may cause further weakening and perturbations of the magnetosphere. 

\section{Acknowledgement}

This work was made possible partly by the satellite monitoring and tracking work of Dr. Jonathan McDowell of Jonathan's Space Report, as well as by the open source data of the Community Coordinated Modeling Center (CCMC). The editing and support of Dr. Samantha Lawler of University of Regina also made this work possible. I'd like to also acknowledge Dr. Andy Lawrence for leading the Starlink FCC appeal, which spurred some of this research.

\vspace{10mm} %10mm vertical space

%\bibliographystyle{plain}}
%\bibliography{mybib}

\begin{thebibliography}{12}
\vspace{10mm} %10mm vertical space
\section{References}

%\bibitem{mag} Anna Anderegg, 2019, Thoughts.

\bibitem{Ionoarticle} Bilitza, D., Pezzopane, M., Truhlik, V., et al, 2022, The International Reference Ionosphere Model: A Review and Description of an Ionospheric Benchmark, Reviews of Geophysics, doi: 10.1029/2022RG000792.

\bibitem{Bolarticle} Boley, A. and Byers, M., 2021, Satellite mega-constellations create risks in Low Earth Orbit, the atmosphere and on Earth, 11, 10642, Scientific Reports, doi: 10.1038/s41598-021-89909-7.

\bibitem{Debyeart}
Champion, K. and Schaub, H., 2022, Effective Debye Lengths in Representative Cislunar Regions, 16th Spacecraft Charging Technology Conference.

\bibitem{https://doi.org/10.1002/ctpp.201900005} Mann, I., Gunnarsdottir, T., Häggström, I. et al, 2019, Radar studies of ionospheric dusty plasma phenomena, doi: 10.1002/ctpp.201900005.

\bibitem{http://planet4589.org/space/con/star/stats.html} McDowell, J. Jonathan's Space Report, 2023, 819, url: http://planet4589.org/space/con/star/stats.html.

\bibitem{http://planet4589.com/space/stats/reentry/data/remass.txt} McDowell, J., Jonathan's Space Report - Reentry masses, 2023, 819, url: http://www.planet4589.com/space/data/reentry/data/remass.txt.

\bibitem{Starlinkmass} McDowell, J., Jonathan's Space Report - Starlink Simulations, 2023, 819, url: https://planet4589.org/astro/starsim/index.html.

\bibitem{Robarticle}
Merlino, R., 2021, Dusty Plasmas: from Saturn's rings to semiconductor processing devices, 6, Advances in Physics, doi: 10.1080/23746149.2021.1873859.

\bibitem{Metalarticle}
Murphy, D., Abou-Ghanem, M., Cziczo, J., et al, 2023, Metals from spacecraft reentry in stratospheric aerosol particles, The Proceedings of the National Academy of Sciences, doi: 10.1073/pnas.2313374120.

\bibitem{mag} Schiermeier, Q., 2013, 
Mission to map Earth's magnetic field readies for take-off, Nature,
doi: 10.1038/nature.2013.14212.

\bibitem{DUSTarticle}
Senel, C., Kaskes, P., Temel, O. et al, 2023, Chicxulub impact winter sustained by fine silicate dust, Nature Geoscience,
doi: 10.1038/s41561-023-01290-4.

\bibitem{mars}
Yokoo, S., Hirose, K., Tagawa, S. and et al, 2022,
Stratification in planetary cores by liquid immiscibility in Fe-S-H, Nature Communications, doi: 10.1038/s41467-022-28274-z.


\end{thebibliography}

%Radar studies of ionospheric dusty plasma phenomena
%I. Mann1 T. Gunnarsdottir1

%http://planet4589.org/space/con/star/stats.html

%http://planet4589.com/space/stats/reentry/data/remass.txt

%Tyler F. M. Brown, Michele T. Bannister & Laura E. Revell (2023) Envisioning a sustainable future for space launches: a review of current research and policy, Journal of the Royal Society of New Zealand, DOI: 10.1080/03036758.2022.2152467

%Boley, A.C., Byers, M. Satellite mega-constellations create risks in Low Earth Orbit, the atmosphere and on Earth. Sci Rep 11, 10642 (2021). https://doi.org/10.1038/s41598-021-89909-7

\end{document}